\documentclass[aps,prl,reprint,preprintnumbers,showkeys,superscriptaddress,floatfix,nofootinbib,nobibnotes]{revtex4-2}

\usepackage{graphicx}
\usepackage{amsmath,amssymb,amsfonts}
\usepackage{mathtools,physics}
\usepackage{booktabs}
\usepackage{hyperref}
\usepackage{cleveref}
\usepackage{orcidlink}

\hypersetup{
pdftitle={Suppressed Direct Detection of Dark Matter from Non-Invertible Selection Rule},
pdfauthor={Riasat Sheikh and Koji Tsumura},
colorlinks=true,
citecolor=blue,
linkcolor=blue,
urlcolor=blue
}

\crefname{equation}{Eq.}{Eqs.}
\Crefname{equation}{Equation}{Equations}
\crefname{figure}{Fig.}{Figs.}
\Crefname{figure}{Figure}{Figures}
\crefname{table}{Table}{Tables}
\Crefname{table}{Table}{Tables}

\begin{document}

\title{Suppressed Direct Detection of Dark Matter \texorpdfstring{\\}{ }from a Non-Invertible Selection Rule}

\author{Riasat Sheikh\,\orcidlink{0009-0007-1207-1358}}
\email[Corresponding author: ]{riasat.sheikh@phys.kyushu-u.ac.jp}
\affiliation{Department of Physics, Kyushu University, 744 Motooka, Nishi-ku, Fukuoka, 819-0395, Japan}

\author{Koji Tsumura\,\orcidlink{0000-0003-3765-2750}}
\email[Corresponding author: ]{tsumura.koji@phys.kyushu-u.ac.jp}
\affiliation{Department of Physics, Kyushu University, 744 Motooka, Nishi-ku, Fukuoka, 819-0395, Japan}
\affiliation{Research Center for Advanced Particle Physics, Kyushu University, 744 Motooka, Nishi-Ku, Fukuoka 819-0395, Japan}
\affiliation{Quantum and Spacetime Research Institute, Kyushu University, 744 Motooka, Nishi-Ku, Fukuoka 819-0395, Japan}
\affiliation{Kavli IPMU (WPI), UTIAS, University of Tokyo, Kashiwa, 277-8584, Japan}

\date{\today}

\begin{abstract}
  Although the existence of dark matter is well established, its particle
  nature is still unknown. Direct detection experiments have been
  struggling to find any viable signal, implying that the dark matter--nucleon
  scattering cross-section must be suppressed. Here we
  propose a mechanism to explain the origin of this suppression by
  considering a non-invertible selection rule. Our numerical analysis
  shows that a suppressed yet predictable dark matter--nucleon scattering cross
  section is possible.
\end{abstract}

\preprint{KYUSHU-HET-385}
\maketitle

\textit{Introduction.}---Astrophysical and cosmological observations
indicate that most of the matter in the Universe is dark matter (DM)
\cite{Bertone:2004pz,Planck:2018vyg}, whose particle nature remains
elusive. Among the many candidates, a weakly interacting particle coupled
to the Standard Model (SM) can reach the observed abundance by thermal
freeze-out. These couplings generically also induce DM--nucleon
scattering. Direct detection experiments have found no viable signal
\cite{PandaX-4T:2021bab,XENON:2023cxc,LZ:2024zvo}, and their limits on the
spin-independent cross-section strongly constrain these couplings
\cite{Arcadi:2017kky,Arcadi:2024ukq}.

The null results can be accommodated if DM annihilates into hidden-sector
states \cite{Pospelov:2007mp}, or if additional annihilation or
coannihilation channels set the abundance \cite{Casas:2017jjg}. These rely
on particular mass relations or choices of parameters. A more robust
suppression arises if the DM is a Nambu-Goldstone boson of a spontaneously
broken global symmetry. Since DM needs a mass, the symmetry must also be
broken explicitly, and if the breaking is only by a mass term then it
becomes a pseudo-Nambu-Goldstone boson (pNGB). It is endowed with
derivative couplings, and thus the DM--nucleon scattering amplitude
mediated by scalars vanishes at tree-level in the zero momentum transfer
limit. DM annihilation, however, remains possible to set the observed
relic abundance \cite{Gross:2017dan}. This mechanism has been realized in
several extensions of the SM
\cite{Karamitros:2019ewv,Abe:2020iph,Okada:2020zxo,Abe:2024vxz}.

Symmetry-breaking quartic couplings, on the other hand, spoil this
mechanism, and if unsuppressed they give a cross-section similar to that
of a conventional Higgs portal \cite{Cline:2013gha}. The smallness of
these couplings has been argued on the basis of a spurion assignment
\cite{Gross:2017dan}. An unbroken ordinary symmetry cannot allow the
symmetry-breaking mass term while forbidding all the symmetry-breaking
quartic couplings.

Thus, the question arises as to whether the mass of DM and the symmetry
breaking quartic couplings can share a common origin that fixes their
relative size. In this Letter we propose such an origin based on a
non-invertible selection rule \cite{Kaidi:2024wio}. The basic idea is
that, if we consider a light sector comprised of SM and beyond-SM (BSM)
fields, and a heavy BSM sector, then the aforementioned rule forbids all
renormalizable breaking among the light fields at tree level but allows
their coupling to a heavy real scalar. The breaking then reaches them
through quantum corrections from the heavy scalar, which generates the mass
and the symmetry-breaking quartic couplings together. These couplings are
then suppressed by the squared ratio of the DM mass to the heavy-scalar
mass, and the DM--nucleon cross-section they induce is suppressed yet
predictable. Such rules and their loop-level violation have been studied
in \cite{Suzuki:2025oov,Suzuki:2025bxg,Suzuki:2025kxz,Dong:2026crl}, and
used to generate neutrino masses \cite{Cordova:2022fhg} and the
$\mathbb{Z}_2$-breaking mass term of the two-Higgs-doublet model
\cite{Hamada:2026sui}.

\textit{Setup.}---We illustrate the mechanism with the two-scalar model
\cite{Abe:2024vxz}, which is the simplest ultraviolet complete realization
of \cite{Gross:2017dan} and has a real, stable DM candidate. The model
contains two complex SM-singlet scalars, $\phi_1^{}$ and $\phi_2^{}$, with
the same unit charge under a gauged ${\rm U}(1)_V$. It also has a global
symmetry, ${\rm U}(1)_a$, under which these two fields rotate by opposite
phases. When the scalars acquire vacuum expectation values, this symmetry
is broken spontaneously and its Goldstone boson is then the DM candidate.
It gains a mass only if ${\rm U}(1)_a$ is also broken explicitly.

Because $\phi_{1,2}^{}$ carry the same charge, every ${\rm U}(1)_a$
breaking term is built from $\phi_1^*\phi_2^{}$ and its conjugate.
Therefore, with an exchange symmetry $\mathbb{Z}_2^{\rm ex}$ (see
\cref{tab:fields}) which stabilizes the DM, the possible renormalizable
operators are
\begin{align}
  V_{\rm br} & = -m_{12}^2\phi_1^*\phi_2^{}
  \nonumber                                                                 \\
             & \qquad
  - \lambda_4(\phi_1^*\phi_2^{})^2
  - \lambda_5(\abs{\phi_1^{}}^2 + \abs{\phi_2^{}}^2)\phi_1^*\phi_2^{}
  \nonumber                                                                 \\
             & \qquad - \lambda_6\abs{H}^2\phi_1^*\phi_2^{} + {\rm H.c.}\,,
  \label{eq:breaking}
\end{align}
where $H$ is the SM Higgs doublet. That model, however, only keeps the
soft-breaking bilinear and omits $\lambda_{4,5,6}$. We add a heavy SM-singlet real
scalar $S$ that couples to the light fields through the operator $S^2\phi_1^*\phi_2^{}$.

Now, following \cite{Kaidi:2024wio}, we assign each field a conjugacy
class (and its conjugate field the inverse class) as
\begin{equation}
  \begin{gathered}
    H \to C_0 = \qty{(0,0,0)},\\
    \phi_1^{} \to C_z = \overline{C}_{\bar z} = \qty{z},\quad
    \phi_2^{} \to C_{\bar z} =  \overline{C}_z = \qty{\bar z},\\
    S \to C_{(0,1,2)} = \overline{C}_{(0,1,2)} = \qty\big{\sigma(0,1,2):\sigma \in S_3},
  \end{gathered}
  \label{eq:classes}
\end{equation}
with $z = (1,1,1)$ and $\bar z = (2,2,2)$. These are classes of the finite group
$G = \Delta(54) = A \rtimes S_3$, where $A$ consists of the triples
$(x_1^{}, x_2^{}, x_3^{})$ of integers mod $3$ with
$\sum_{i=1}^3 x_i^{} = 0$, added componentwise, and $S_3$ permutes the
entries. (This is the smallest group
we find that can realize the mechanism, and a systematic search will be presented in
a subsequent work.) Conjugation acts on the
abelian $A$ only through these permutations, so the classes inside $A$
are its $S_3$ orbits, and $C_0$, $C_z$ and $C_{\bar z}$ form the center
of $G$ \cite{Escobar:2008vc,Ishimori:2010au}.  The field assignments are summarized in \cref{tab:fields}.
A coupling is allowed at tree level when the product of the classes of its
fields contains the neutral identity $C_0$. A loop can generate the
forbidden couplings, since the product of two classes generally contains
several classes.

\begin{table}[t]
  \centering
  \setlength{\tabcolsep}{0.5em}
  \renewcommand{\arraystretch}{1.12}
  \begin{tabular}{lccccc}
    \toprule
    field       & SM                                   & ${\rm U}(1)_V$ & ${\rm U}(1)_a$ & $\mathbb{Z}_2^{\rm ex}$ & class         \\
    \midrule
    $\phi_1^{}$ & $(\mathbf{1}, \mathbf{1}, 0)$        & $+1$           & $+1$           & $\phi_2^{}$             & $C_z$         \\
    $\phi_2^{}$ & $(\mathbf{1}, \mathbf{1}, 0)$        & $+1$           & $-1$           & $\phi_1^{}$             & $C_{\bar z}$  \\
    $S$         & $(\mathbf{1}, \mathbf{1}, 0)$        & $0$            & $0$            & $S$                     & $C_{(0,1,2)}$ \\
    $H$         & $(\mathbf{1}, \mathbf{2}, \tfrac12)$ & $0$            & $0$            & $H$                     & $C_0$         \\
    \bottomrule
  \end{tabular}
  \caption{Field content and class assignments. The
  $\mathbb{Z}_2^{\rm ex}$ column gives the image of each field under the
  exchange. All other SM fields are $\mathrm{U}(1)_V$ neutral and
  carry identity $C_0$.}
  \label{tab:fields}
\end{table}

Therefore, in \cref{eq:breaking}, although $\abs{\phi_{1,2}^{}}^2$ and
$\abs{H}^2$ are neutral, the bilinear $\phi_1^*\phi_2^{}$ carries only
$C_{\bar z}\cdot C_{\bar z} = C_z$, so every term is forbidden at tree
level. On the other hand, for the heavy scalar, $S^2$ contains the classes
of the light fields along with the identity,
\begin{equation}
  C_{(0,1,2)} \cdot C_{(0,1,2)}
  = 6\,\qty\big(C_0 + C_z + C_{\bar z}) + 3\,C_{(0,1,2)}\,,
  \label{eq:key}
\end{equation}
so it allows the $S^2\phi_1^*\phi_2^{}$ operator while the bilinear alone
stays forbidden.
Symmetry-breaking expectation values or additional spurions can evade
this, but the relative size of the symmetry-breaking mass and quartic
couplings is then set by the spurion assignment rather than fixed by the
DM mass. Part of the rule survives at all orders as an ordinary
$\mathbb{Z}_2$ symmetry \cite{Kaidi:2024wio,Suzuki:2025kxz}, and all the
fields in \cref{eq:classes} are neutral under it.

The rule is also compatible with the exchange symmetry $\mathbb{Z}_2^{\rm
ex}$, since the automorphism of $G$ that interchanges $C_z$ and $C_{\bar
z}$ preserves $C_{(0,1,2)}$. Therefore, the most general renormalizable
potential allowed by the rule that respects the exchange symmetry is
\begin{align}
  V            & = V_0 + V_{\rm br}^S,
  \nonumber                                                                                                      \\
  V_0          & = \mu_H^2\,\abs{H}^2 + \lambda_H^{}\,\abs{H}^4
  + \lambda_{H\phi}^{}\,\abs{H}^2\qty(\abs{\phi_1^{}}^2 + \abs{\phi_2^{}}^2)
  \nonumber                                                                                                      \\
               & \quad\quad + \mu_\phi^2\,\qty(\abs{\phi_1^{}}^2 + \abs{\phi_2^{}}^2)
  + \lambda_\phi\,\qty(\abs{\phi_1^{}}^4 + \abs{\phi_2^{}}^4)
  \nonumber                                                                                                      \\
               & \quad\quad + \lambda_{12}^{}\,\abs{\phi_1^{}}^2\abs{\phi_2^{}}^2
  + \frac{\mu_S^2}{2}S^2 + \frac{\mu_3^{}}{3!}S^3 + \frac{\lambda_{S}^{}}{4!}S^4
  \nonumber                                                                                                      \\
               & \quad\quad + \frac12S^2\qty\Big[\lambda_{S\phi}^{}\,\qty(\abs{\phi_1^{}}^2 + \abs{\phi_2^{}}^2)
  + \lambda_{HS}^{}\,\abs{H}^2],
  \nonumber                                                                                                      \\
  V_{\rm br}^S & = \frac{\lambda_{S\phi}'}{2}\,S^2\qty\big(\phi_1^*\phi_2^{} + {\rm H.c.}),
  \label{eq:pot}
\end{align}
where all couplings are real without any loss of generality and $V_{\rm br}^S$ is the only renormalizable source of ${\rm U}(1)_a$ breaking.
The rule forbids the terms linear in $S$, namely $S$, $S\abs{H}^2$,
$S\abs{\phi_i^{}}^2$ and $S\phi_1^*\phi_2^{}$, so $\expval{S} = 0$ at tree
level. (A detailed study will be presented in the subsequent work.)

\textit{Radiative hierarchy.}---We now integrate out $S$ at one-loop
order. In a background of the light fields, $S$ has the mass $m_S^2 =
\eval{\partial^2V/\partial S^2}_{S=0}$ with $V$ of \cref{eq:pot}, and its
one-loop Coleman--Weinberg potential at the renormalization scale
$\Lambda$ is
\begin{equation}
  V_S = \frac{m_S^4}{64\pi^2}
  \qty(\ln\frac{m_S^2}{\Lambda^2} - \frac32).
  \label{eq:CW}
\end{equation}
By imposing the rule at the scale $\Lambda$, we ensure the couplings of
\cref{eq:breaking} vanish there in the $\overline{\rm MS}$ scheme,
and we neglect any finite threshold corrections from the ultraviolet completion.
We expand the light fields about the exchange-preserving
vacuum $\expval{\phi_1^{}} = \expval{\phi_2^{}} = v_\phi^{}/2$ and
$\expval{H} = \qty(0,v_h^{}/\sqrt{2})^{\rm T}$, with
$v_h^{} = 1/\sqrt{\mathstrut\sqrt2 G_F^{}}\approx 246~{\rm GeV}$,
where $G_F^{}$ is the Fermi constant.
In the unitarity gauge, the light fields read
\begin{equation}
  \begin{gathered}
    \phi_1^{} = \frac12\qty(v_\phi^{} + s_+^{} + s_-^{} + i a),\\
    \phi_2^{} = \frac12\qty(v_\phi^{} + s_+^{} - s_-^{} - i a),\\
    H = \frac{1}{\sqrt{2}}\pmqty{0\\v_h^{} + h},
  \end{gathered}
\end{equation}
where $s_+^{}$ and $h$ are exchange even and $s_-^{}$ and $a$ exchange
odd, so the DM candidate $a$ is stable when it is lighter than $s_-^{}$.
In this vacuum configuration, $S$ has the mass
\begin{equation}
  M_S^2 = \mu_S^2 +
  \frac12 \qty\Big[\qty(\lambda_{S\phi}^{} + \lambda_{S\phi}')\, v_\phi^2 + \lambda_{HS}^{}\,v_h^2],
  \label{eq:MS}
\end{equation}
which receives a fixed contribution from the electroweak symmetry-breaking
scale $v_h^{}$, but is naturally driven by the dark symmetry-breaking scale
$v_\phi^{}$ through $\lambda_{S\phi}^{}$ and $\lambda_{S\phi}'$. After imposing the
vacuum stationarity conditions, and defining
$\ell \equiv \ln(\Lambda^2/M_S^2)$, \cref{eq:CW} gives the DM mass as
\begin{equation}
  m_{\rm DM}^2
  = \frac{\lambda_{S\phi}'}{16\pi^2}\,M_S^2\,\qty(\ell + 1)\,,
  \label{eq:mass}
\end{equation}
so that $\lambda_{S\phi}' = 16\pi^2\varepsilon/\qty(\ell + 1)$, where
$\varepsilon \equiv \flatfrac{m_{\rm DM}^2}{M_S^2}$ is the squared ratio of
the DM mass to the heavy-scalar mass.
Thus, the DM mass is generated by the heavy sector and lies below $M_S$ by
a loop factor. Moreover, since $\lambda_{S\phi}'$ is the only coupling
that breaks ${\rm U}(1)_a$, the limit $\varepsilon \to 0$ ($M_S \to
  \infty$ for a fixed DM mass) restores the symmetry. Hence, we refer to
$\varepsilon$ as the pNGB quality parameter. The quadratic terms of the
expansion of \cref{eq:CW} in $m_S^2 - M_S^2$ about the vacuum then give
the symmetry-breaking quartic couplings of \cref{eq:breaking},
\begin{equation}
  \lambda_4 = \frac{\kappa}{4}\,\varepsilon\,\lambda_{S\phi}'\,,\quad
  \lambda_5 = \frac{\kappa}{2}\,\varepsilon\,\lambda_{S\phi}^{}\,,\quad
  \lambda_6 = \frac{\kappa}{2}\,\varepsilon\,\lambda_{HS}^{}\,,
  \label{eq:hard}
\end{equation}
with $\kappa \equiv \ell/(\ell + 1)$, close to unity for $\Lambda \gg M_S$.
Thus, the symmetry-breaking quartic couplings are suppressed by the pNGB quality parameter.

\textit{Direct detection prospects.}---We now focus on the DM--nucleon
scattering. The usual cancellation mechanism (see \cite{Gross:2017dan})
that suppresses the scattering in the zero momentum transfer limit still
holds. In our case, the symmetry-breaking quartic couplings induce
additional DM interactions with exchange-even mediators $s_+^{}$ and $h$,
\begin{align}
  \mathcal L
   & \supset -\qty\Big[\qty\big(2\lambda_4 + \lambda_5)v_\phi^{}s_+^{} + \lambda_6v_h^{}h]a^2
  \nonumber                                                                                   \\
   & \quad = -\frac{\kappa}{2}\,\varepsilon\,
  \qty\Big[\qty\big(\lambda_{S\phi}^{} + \lambda_{S\phi}')v_\phi^{}s_+^{} + \lambda_{HS}^{}v_h^{}h]a^2\,,
  \label{eq:hierarchy}
\end{align}
which opens up a possibility to probe the DM in direct detection
experiments.

The exchange-even mediators mix into the mass eigenstates $h_1 =
h\cos\theta + s_+^{}\sin\theta$ (the observed Higgs boson) and $h_2 =
-h\sin\theta + s_+^{}\cos\theta$, with mixing angle $\theta$ and their
respective masses $m_{h_1}^{} = 125~{\rm GeV}$ and $m_{h_2}^{}$. The SM
fermions couple to them as $\mathcal L \supset -(h_1\cos\theta -
h_2\sin\theta)\sum_f(m_f^{}/v_h^{})\bar ff$, so at zero momentum transfer
the $h_1$ and $h_2$ exchanges give the DM--fermion amplitude $\mathcal M
\propto \lambda_{\rm eff}\, m_f^{}/m_{h_1}^2$, where both exchanges
combine into the effective Higgs-portal coupling
\begin{equation}
  \begin{split}
    \lambda_{\rm eff}
     & = \kappa\,\varepsilon\,
    \Biggl\{
    \lambda_{HS}^{}\qty[1 - \qty(1 - \frac{m_{h_1}^2}{m_{h_2}^2})\sin^2\theta]
    \\
     & \qquad +
    \qty\big(\lambda_{S\phi}^{} + \lambda_{S\phi}')
    \qty(\frac{v_\phi^{}}{v_h^{}})
    \qty(1 - \frac{m_{h_1}^2}{m_{h_2}^2})\sin\theta\cos\theta
    \Biggr\}\,.
  \end{split}
  \label{eq:gdef}
\end{equation}
Therefore, the non-relativistic spin-independent DM--nucleon elastic cross-section can be written as
\begin{equation}
  \sigma_{\rm SI}^{} = \frac{f_N^2}{4\pi}\,
  \frac{\mu_N^2m_N^2}{m_{\rm DM}^2m_{h_1}^4}\,\lambda_{\rm eff}^2\,,
  \label{eq:sigma}
\end{equation}
where $m_N^{}$ is the nucleon mass, $\mu_N^{}$ the DM--nucleon reduced mass,
and $f_N^{} \simeq 0.3$. The cross-section thus vanishes
in the symmetric limit ($\varepsilon \to 0$), and at fixed $M_S$ it grows with the DM mass, unlike a
Higgs portal with a fixed coupling.

As a benchmark, we fix $\lambda_{S\phi}^{} = \lambda_{S\phi}'$ (a
sufficient choice for boundedness from below), $\lambda_{HS}^{} = 1$,
$\Lambda = 100M_S$, $m_{h_2}^{} = 300~{\rm GeV}$, $\abs{\sin\theta} =
0.1$, the ${\rm U}(1)_V$ gauge-boson mass $m_{Z'}^{} = 200~{\rm GeV}$, and
$m_{s_-}^{} = 1.5m_{\rm DM}^{}$. We take a gauge kinetic mixing angle of
$10^{-4}$ between ${\rm U}(1)_V$ and hypercharge, allowing $Z'$ to decay
into SM particles. We use \texttt{FeynRules} \cite{Alloul:2013bka} to
generate the model files, including the induced couplings of
\cref{eq:hard}, and solve the Boltzmann equation for $a$ with
\texttt{micrOMEGAs} \cite{Alguero:2023zol} to fix the dark symmetry
breaking scale $v_\phi^{}$ by the observed relic abundance, $\Omega h^2 =
0.120 \pm 0.001$ \cite{Planck:2018vyg}.

\begin{figure*}[t]
  \includegraphics[width=\textwidth]{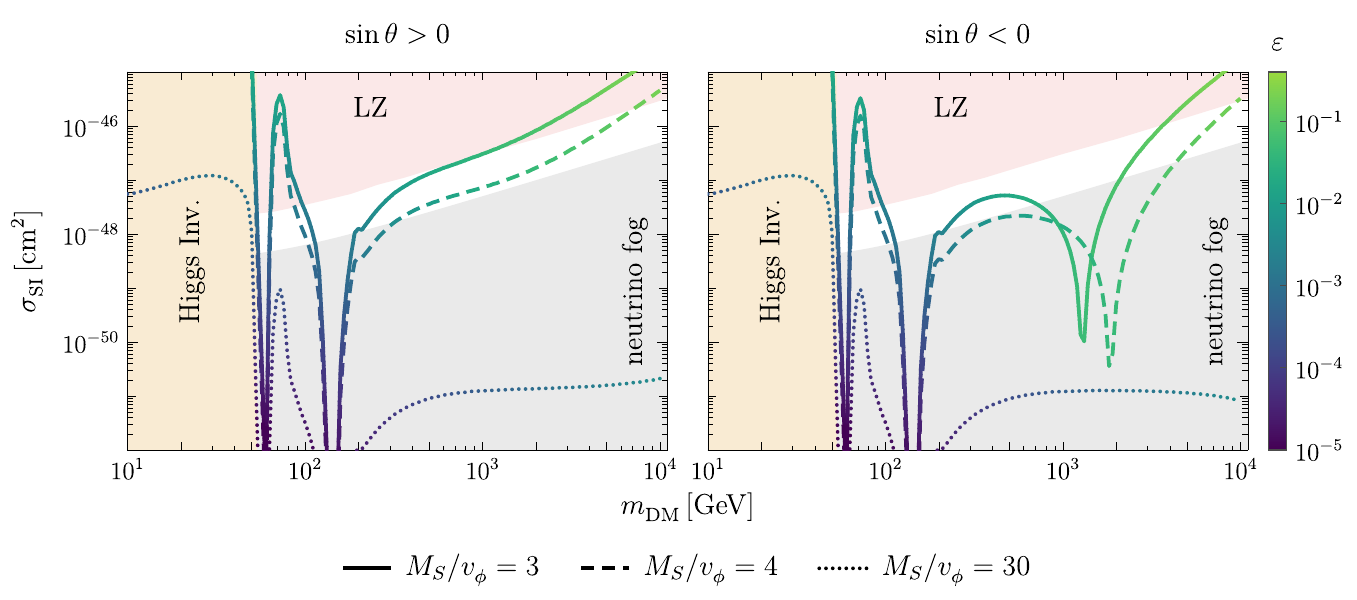}
  \caption{Spin-independent DM--nucleon elastic cross-section. The curves
  are colored by the pNGB quality parameter $\varepsilon$. The red
  region is excluded by LZ, the yellow region by invisible Higgs decays, and
  the gray region is the neutrino fog. The
  left (right) panel is for positive (negative) scalar mixing angle $\sin\theta$.}
  \label{fig:sigma}
\end{figure*}

In \cref{fig:sigma} we show the cross-section of \cref{eq:sigma} along the
relic curve $v_\phi^{}(m_{\rm DM}^{})$ for fixed ratios $M_S/v_\phi^{}$.
The left (right) panel is for positive (negative) scalar mixing angle
$\sin\theta$. The red and yellow regions are excluded by the LZ experiment
limit \cite{LZ:2024zvo} and the Higgs invisible decay bound
\cite{ATLAS:2023tkt,CMS:2023sdw}, respectively. The gray region is the
neutrino fog \cite{OHare:2021utq}, where the DM becomes statistically
difficult to separate from the background neutrinos in a detector. The
curves are colored by the pNGB quality parameter $\varepsilon = (m_{\rm
DM}^{}/v_\phi^{})^2(v_\phi^{}/M_S)^2$. With the second factor fixed, the
relic curve maps directly onto the pNGB quality. Near the $m_{\rm DM}^{}
\simeq m_{h_{1,2}}^{}/2$ resonance dips, DM annihilation is efficient, and
the observed abundance requires a large $v_\phi^{}$, hence the small
$\varepsilon$; elsewhere $\varepsilon$ grows with the DM mass.

The window between the LZ limit and the fog spans less than an order of
magnitude in $\sigma_{\rm SI}^{}$, and therefore selects a narrow range of
$\varepsilon$. For $m_{\rm DM}^{} < m_{Z'}^{}$, we have $\varepsilon
\lesssim 10^{-3}$ for both signs of the mixing, and above it $\varepsilon$
grows. For negative mixing, however, we have an accidental cancellation.
The two terms of \cref{eq:gdef} interfere destructively and move the
window to smaller $M_S/v_\phi^{}$ even with poor pNGB quality, and near
the zeros of $\lambda_{\rm eff}$, hide the DM below the fog. Moreover, for
$M_S/v_\phi^{} = 30$ the DM lies deep in the fog, approaching the softly
broken model \cite{Abe:2024vxz}. Thus suppressed yet predictable
DM--nucleon scattering is possible.

\textit{Summary.}---We have presented a mechanism that can explain the
origin of the soft breaking of a global symmetry that gives mass to a pNGB
DM candidate, while suppressing the symmetry-breaking quartic couplings
that would otherwise induce a large DM--nucleon scattering cross-section.
The mechanism relies on a non-invertible selection rule that forbids these
couplings at tree level, while allowing them to be generated at one loop
through the coupling of the light fields to a heavy scalar. This results
in a natural hierarchy between the symmetry-breaking mass terms and
quartic couplings, controlled by the pNGB quality parameter $\varepsilon$.
Our numerical analysis shows that a suppressed yet predictable DM--nucleon
scattering cross-section is possible. A detailed analysis will be
presented in a subsequent work.

\textit{Acknowledgments.}---This work was supported in part by JSPS
Grant-in-Aid for Scientific Research KAKENHI Grants No. JP22K03620 (K.T.).
This work was supported by the Sasakawa Scientific Research Grant from The
Japan Science Society (R.S.).

\bibliography{references}
\end{document}